\documentclass[aps,prb,twocolumn,amssymb,amsmath,superscriptaddress]{revtex4-1}
\usepackage{bm}
\usepackage{times}
\usepackage{graphicx}
\usepackage{color}
\usepackage{dcolumn}
\usepackage[colorlinks=true, letterpaper=true, pdfstartview=FitV, linkcolor=blue, citecolor=blue, urlcolor=blue]{hyperref}
\usepackage{appendix}

\begin{document}

\title{Spin separation and filtering assisted by topological corner states in the Kekul\'{e} lattice}

\author{Kai-Tong Wang}
%\email[]{ktwang@haust.edu.cn}
\affiliation{School of Physics and Engineering, Henan University of Science and Technology, Luoyang 471023, China}

\author{Hui Wang}
\affiliation{School of Physics and Engineering, Henan University of Science and Technology, Luoyang 471023, China}

\author{Shijie Liu}
\affiliation{School of Physics and Engineering, Henan University of Science and Technology, Luoyang 471023, China}

\author{Miaomiao Wei}
\affiliation{College of Physics and Optoelectronic Engineering, Shenzhen University, Shenzhen 518060, China}

\author{Fuming Xu}
\email[]{xufuming@szu.edu.cn}
\affiliation{College of Physics and Optoelectronic Engineering, Shenzhen University, Shenzhen 518060, China}
\affiliation{Quantum Science Center of Guangdong-Hongkong-Macao Greater Bay Area (Guangdong), Shenzhen 518045, China}

\begin{abstract}
Higher-order topological corner states have been realized in two-dimensional Kekul\'{e} lattice, which can be further coupled with spin polarization through the implementation of local magnetization. In this work, we numerically investigate the spin-dependent transport properties assisted by topological corner states in the Kekul\'{e} lattice. By applying local magnetization and electric potential, the topological corner states are spin polarized with opposite spins localized at different corners, thereby demonstrating a spin-corner state locking mechanism. Transport characteristics, including transmission, local density of states, and local current density, are calculated for a two-terminal setup consisting of a diamond-shaped Kekul\'{e} lattice connected to two leads. When opposite local magnetization is applied to the corners, spin-up and spin-down electrons are perfectly separated, forming two spin-polarized conducting channels and leading to spin spatial separation. In the presence of identical local magnetization on both corners and an electric potential at one corner, the spin-polarized corner states can facilitate selective filtering of different spins and generate spin-polarized currents by tuning the energy. Furthermore, spin-resolved transmission diagrams as functions of both the Fermi energy and electric potential are presented, illustrating the global distribution of spin filtering through topological corner states.
\end{abstract}

\maketitle

\section{INTRODUCTION}

Topological insulators (TIs) exhibit unique electronic properties crucial for energy-efficient electronics and spintronics. Recently, higher-order topological insulators(HOTIs) have attracted significant research interest.\cite{ref1,ref2,ref3,ref4,ref43,ref44,ref45,Guo2023} HOTIs are characterized by symmetry-protected topological corner states and hinge states in two- or three-dimensional materials.\cite{ref6,ref7,ref8,Sheng2020,ref9,ref10,ref46,Sun2022,Hasan23} So far, HOTIs have been theoretically predicted and experimentally realized in various systems, including the kagome lattice,\cite{ref11,ref12} bismuth crystal,\cite{ref13,ref14} twisted bilayer graphene,\cite{ref15,ref16} transition metal dichalcogenides,\cite{ref17,ref18} and diverse metamaterials.\cite{ref19,ref20,ref21,ref22,ref23,ref24,ref47,ref48,Sun2023} Notably, topological corner states feature fractional charge and a zero-dimensional localized nature, offering a potential platform for quantum calculation,\cite{ref25} information processing,\cite{ref26} and the design of optical cavity modes.\cite{ref27}

%Recently, quantum transport response of topological hinge modes in $\alpha$-Bi$_4$Br$_4$ has been numerically revealed through prominent Aharonov-Bohm oscillations.\cite{Hasan23}

Previous studies have shown that internal degrees of freedom in materials can be coupled with topological corner states. For instance, valley-polarized corner states were discussed in higher-order topological sonic crystals,\cite{ref28} while pseudospin-polarized corner states were proposed in topological quadrupole insulators.\cite{ref29} The real spin degree of freedom can also be engineered with higher-order topological phases. It was experimentally confirmed that topological corner states can be generated by tuning the intracell and intercell hopping parameters in a two-dimensional honeycomb lattice with Kekul\'{e}-like hopping textures.\cite{ref30} Furthermore, by introducing local magnetization and electric potential in the Kekul\'{e} lattice, it has been demonstrated that the spin can be manipulated,\cite{ref31} leading to spin-selective topological corner states. To explore the potential application of topological corner states in spintronics, it is necessary to investigate spin-polarized transport properties associated with them.

Unlike the conducting edge or surface states of TIs and hinge modes of HOTIs,\cite{Hasan23} topological corner states are localized in real space and isolated within an energy gap. Consequently, investigating the transport properties of topological corner states is challenging, with only a few exceptions. The resonant tunneling behavior related to charge transport of topological corner states were addressed in two-dimensional honeycomb lattices.\cite{ref32,Xu2022} It was also proposed that quantized pure spin currents can be generated in a variety of second-order TIs through the spin pumping mechanism.\cite{ref5} Multi-terminal nano-switches based on topological corner states was demonstrated to be robust against disorder scattering.\cite{ref51} However, it remains intriguing to explore how spin-polarized topological corner states influence the transport properties in the Kekul\'{e} lattice.

In this work, we numerically investigate the spin-dependent transport assisted by topological corner states in the Kekul\'{e} lattice. A two-terminal transport setup is considered, which consists of a diamond-shaped Kekul\'{e} flake surrounded by armchair boundaries and connected to two conducting leads. The transmission, local density of states, and local current density are calculated in detail. In a pristine Kekul\'{e} lattice without local magnetization, it is found that the resonant tunneling channels are builded, when incident electrons transform the localized corner states into extended states across the central scattering region. In the presence of opposite local magnetization at two corners, spin-up and spin-down electrons are locked to different corners, leading to the spin spatial separation. In this case, the resonant tunneling is mediated by spin-polarized corner states and both spin components contribute equally to the total transmission. When identical local magnetization applied to both corners and a local electric potential applied to one corner, spin-polarized corner states can facilitate selective spin filtering and generate spin-polarized currents via tuning the Fermi energy. We also demonstrate spin-resolved transmission diagrams as functions of both the Ferimi energy and electric potential, clearly illustrating the role played by topological corner states in spin filtering.

This paper is organized as follows. In Sec.~\uppercase\expandafter{\romannumeral2}, the model Hamiltonian for the Kekul\'{e} lattice hosting topological corner states as well as the transport formalism are introduced. In Sec.~\uppercase\expandafter{\romannumeral3}, we present the numerical results and relevant discussion. Finally, a brief summary is given in Sec.~\uppercase\expandafter{\romannumeral4}.

\begin{figure}[t]
\centering
\includegraphics[width=\columnwidth]{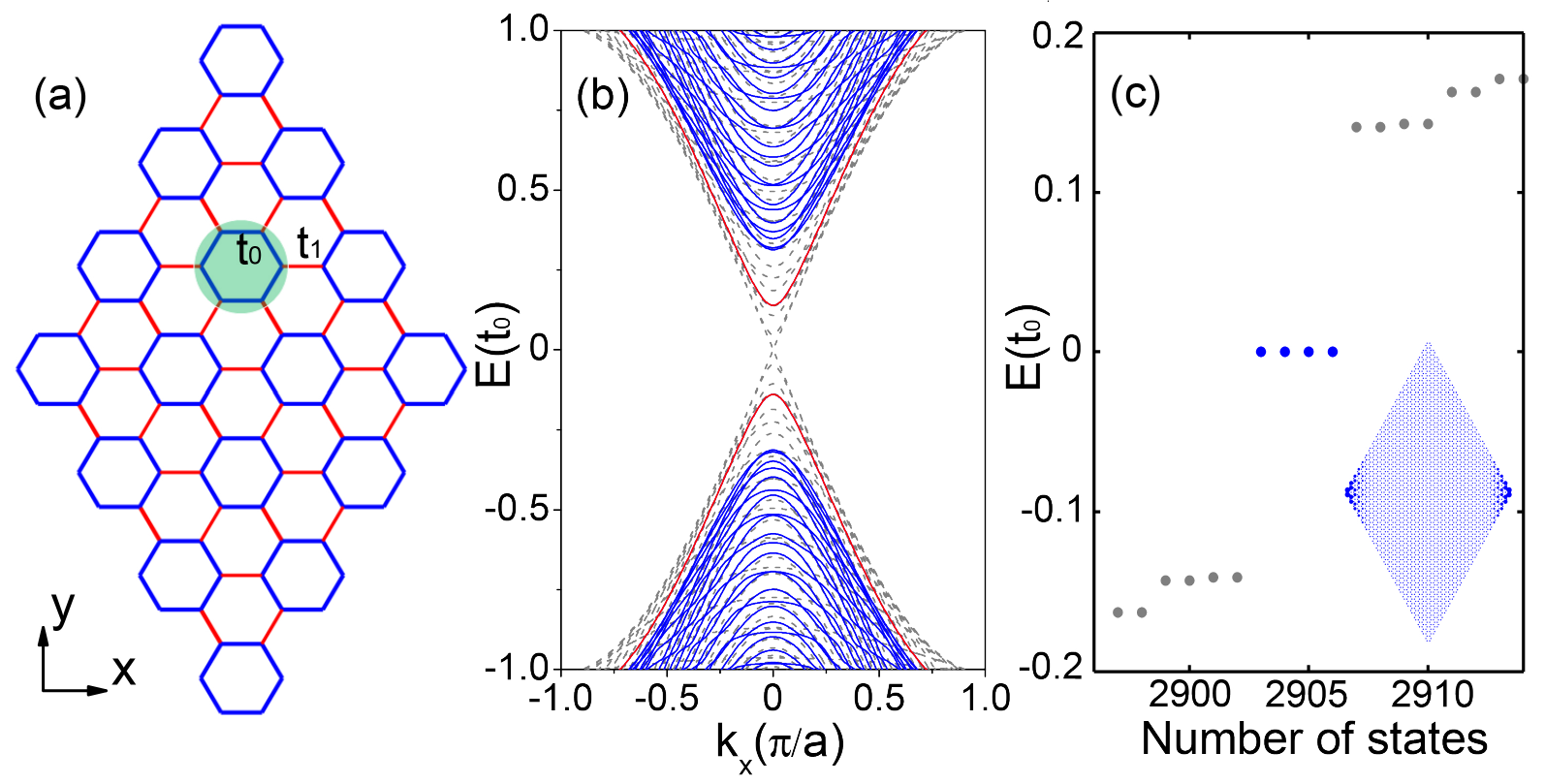}
\caption{(Color online) (a) Schematic of the Kekul\'{e} lattice with intracell hopping $t_{0}$ and intercell hopping $t_{1}$, which are denoted by the blue and red bonds, respectively. The unit cell is highlighted in green circle. (b) Energy bands of a Kekul\'{e} lattice ribbon with armchair boundaries for $t_{1}=1.3 t_{0}$, where the blue and red bands represent the bulk and edge modes, respectively. The ribbon width is $22$ unit cells. The bands of a semimetal armchair ribbon with $t_{1} \approx t_{0}$ are shown as the gray background. (c) Discrete energy levels for an isolated diamond-shaped Kekul\'{e} lattice surrounded by armchair boundaries. The lattice size is $22 \times 22$ unit cells and $t_{1}=1.3 t_{0}$. Inset: the eigenfunction distribution for the zero-energy states labeled by blue circles, which are sharply localized at two corners. }\label{fig1}
\end{figure}

\section{MODEL AND FORMALISM}

We start with the tight-binding Hamiltonian of the Kekul\'{e} lattice, which is expressed as\cite{ref33,ref34}
\begin{equation}
  H_{K} = t_{0}\sum_{<i,j>}d_{i}^{\dag}d_{j} + t_{1}\sum_{<i^{'},j^{'}>}d_{i^{'}}^{\dag}d_{j^{'}},
\end{equation}
where $d_{i}^{\dag}=(d_{i\uparrow}^{\dag},d_{i\downarrow}^{\dag})^{T}$ is the creation operator for an electron with spin up and spin down at site $i$. $t_{0}$ and $t_{1}$ correspond to the intracell and intercell hopping parameters, which are shown by the blue and red bonds in Fig.~\ref{fig1}(a). The unit cell is highlighted in green circle.

The electronic states of the Kekul\'{e} lattice can be effectively modulated by $t_0$ and $t_1$. For an armchair Kekul\'{e} ribbon with the width of $22$ unit cells, a certain band gap exists for $t_{1} < t_0$. With the increasing of $t_{1}$, the band gap gradually closes and the system transforms into a semimetal phase, as shown by the gray bands in Fig.~\ref{fig1}(b) for $t_{1} \approx t_{0}$.\cite{note0} When $t_{1}$ further increases, a topological phase transition occurs with the emergence of helical edge states shown by the red lines in Fig.~\ref{fig1}(b). Previous studies found that,\cite{Duan2017,ref29,ref30,ref34} when $t_{1} > t_{0}$, higher-order topological corner states appear in the Kekul\'{e} lattice. Our numerical results also demonstrate that, topological corner states arise in the diamond-shaped Kekul\'{e} flake surrounded by armchair boundaries, as shown in Fig.~\ref{fig1}(c). Notice that topological corner states reside near the zero energy.

\begin{figure}[tbp]
\centering
\includegraphics[width=\columnwidth]{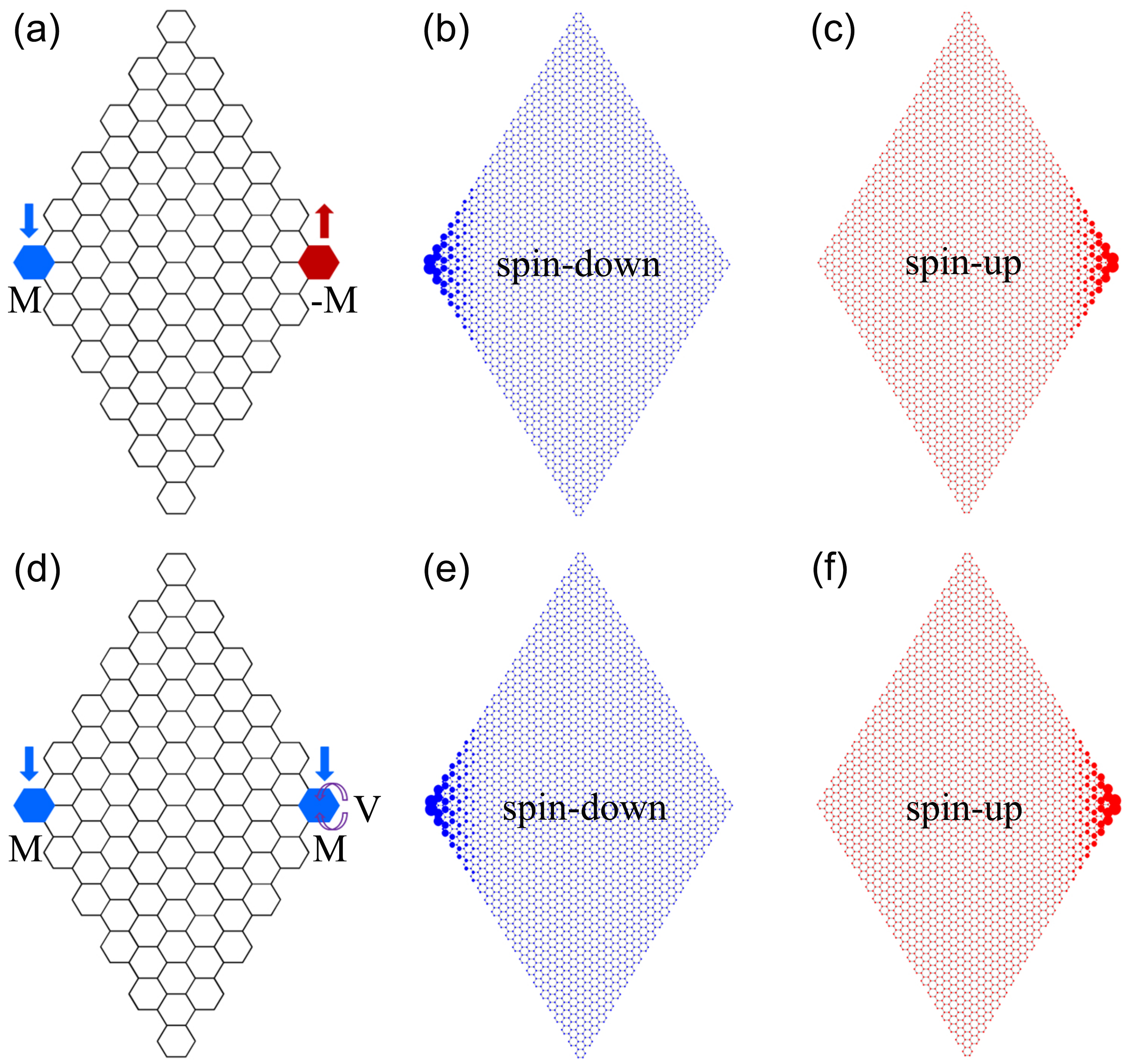}
\caption{(Color online) (a) The diamond-shaped Kekul\'{e} lattice with opposite local magnetization $\pm M$ applied on the left and right corners. (b) and (c): Eigenfunction distributions of the spin-up and spin-down polarized corner states are shown in red and blue, respectively. (d) The Kekul\'{e} lattice with the same local magnetization on both corners and a local electric potential $V$ at the right corner. (e) and (f): Eigenfunction distributions of the spin-polarized corner states. Other parameters: $t_{1}=1.3 t_{0}$, $\lambda=0.3$, and $V=0.4$.}\label{fig2}
\end{figure}

Recently, it was found that by introducing local magnetization on the appropriate corner of the Kekul\'{e} lattice, its topological corner states can be spin-polarized; by further applying an electric potential on this corner, the spin polarization at both corners can be effectively manipulated, realizing the remote control of spin polarization of topological corner states.\cite{ref31} To implement this mechanism, the following Hamiltonian is considered\cite{ref31}
\begin{equation}
H=H_{K}+\lambda \sum_{<i,L>}d_{i}^{\dag}\sigma_{z} d_{i} + \lambda \nu \sum_{<i,R>} d_{i}^{\dag}\sigma_{z} d_{i} + V\sum_{<i,R>}d_{i}^{\dag} d_{i},
\end{equation}
where $\lambda$ and $V$ denote the exchange field strength and the electric potential, respectively. The local magnetization and the electric potential are introduced on the left(L) and right(R) corners of the diamond-shaped Kekul\'{e} lattice. $\sigma_{z}$ is the Pauli matrix. $\nu=\pm 1$ corresponds to that either the same or opposite local magnetization is applied on the two corners, as shown in Fig.~\ref{fig2}(a) and (d). The electric potential $V$ is only added on the right corner.

The influence of local magnetization and electric potential on topological corner states is illustrated in Fig.~\ref{fig2}. In Fig.~\ref{fig2}(a), opposite local magnetization $\pm M$ is introduced at the left and right corners with obtuse angles, where $\pm M$ is highlighted in the blue and red region. Consequently, the exchange splitting interaction can induce spin polarization of the corner states. As displayed in Fig.~\ref{fig2}(b) and (c), the eigenfunction distributions of the spin-up (red circles) and spin-down (blue circles) states are spatially separated and localized at the right and left corners, respectively. When the same local magnetization is applied on both corners, the resulting corner states are spin degenerate. Interestingly, while further applying a local electric potential $V$ at the right corner (Fig.~\ref{fig2}), spin-flipping process occurs\cite{ref31} and the eigenfunction distributions in Fig.~\ref{fig2}(e) and (f) are clearly spin-polarized. Apparently, spin-polarization topological corner states can be easily achieved by applying local magnetization as well as electric potential in the Kekul\'{e} lattice.

To study the spin-dependent transport properties of topological corner states, the diamond-shaped Kekul\'{e} flake hosting corner states is connected to two conducting leads. The two-terminal transport setup is depicted in Fig.~\ref{fig3}(a). Based on the Landauer-B$\rm \ddot{u}$ttiker formula, the transmission from lead $p$ to lead $q$ is expressed as\cite{ref35}
\begin{equation}\label{}
  T_{p q} = {\rm Tr} [\Gamma_{p}G^{r}\Gamma_{q}G^{a}],
\end{equation}
where $ G^{r}$ ($ G^{a}$) is the retarded (advanced) Green's function, $G^{r} = (E-H_{C}-\Sigma^{r})^{-1}$ and $G^{a} = G^{r, \dagger}$. $H_{C}$ is the Hamiltonian of the central scattering region. $\Sigma^{r}$ is sum over the self-energies from two leads, which can be calculated using the transfer-matrix method.\cite{ref36,ref37} $\Gamma_{p/q}=i(\Sigma^{r}-\Sigma^{a})_{p/q}$ is the linewidth function denoting the interaction between the central region and leads.

The partial local density of states(PLDOS) can dynamically reveal the distribution of a scattering state through the central region, which has the following form\cite{ref38,ref39,ref40}
\begin{equation}\label{}
  \rho_{p}(E) = \frac{1}{2\pi}[G^{r}\Gamma_{p}G^{a}]_{ii}.
\end{equation}
Here $i$ represents the real-space lattice. $\rho_{p}$ corresponds to the electrons incident from lead $p$ and dwelling inside the scattering region. On the other hand, the local current density provides more subtle insight into the transport details. The local current from site $i$ to $j$ can be calculated as\cite{ref41,ref42}
\begin{equation}\label{}
  J_{i\rightarrow j} = \frac{2e^2}{h} {\rm Im} [\sum_{\alpha,\beta}H_{i\alpha,j\beta} (G^{r}\Gamma_{p}G^{a})_{j\beta,i\alpha}],
\end{equation}
with $\alpha$/$\beta$ denoting the spin index.

\section{NUMERICAL RESULTS AND DISCUSSION}

\begin{figure}[tbp]
\includegraphics[width=\columnwidth]{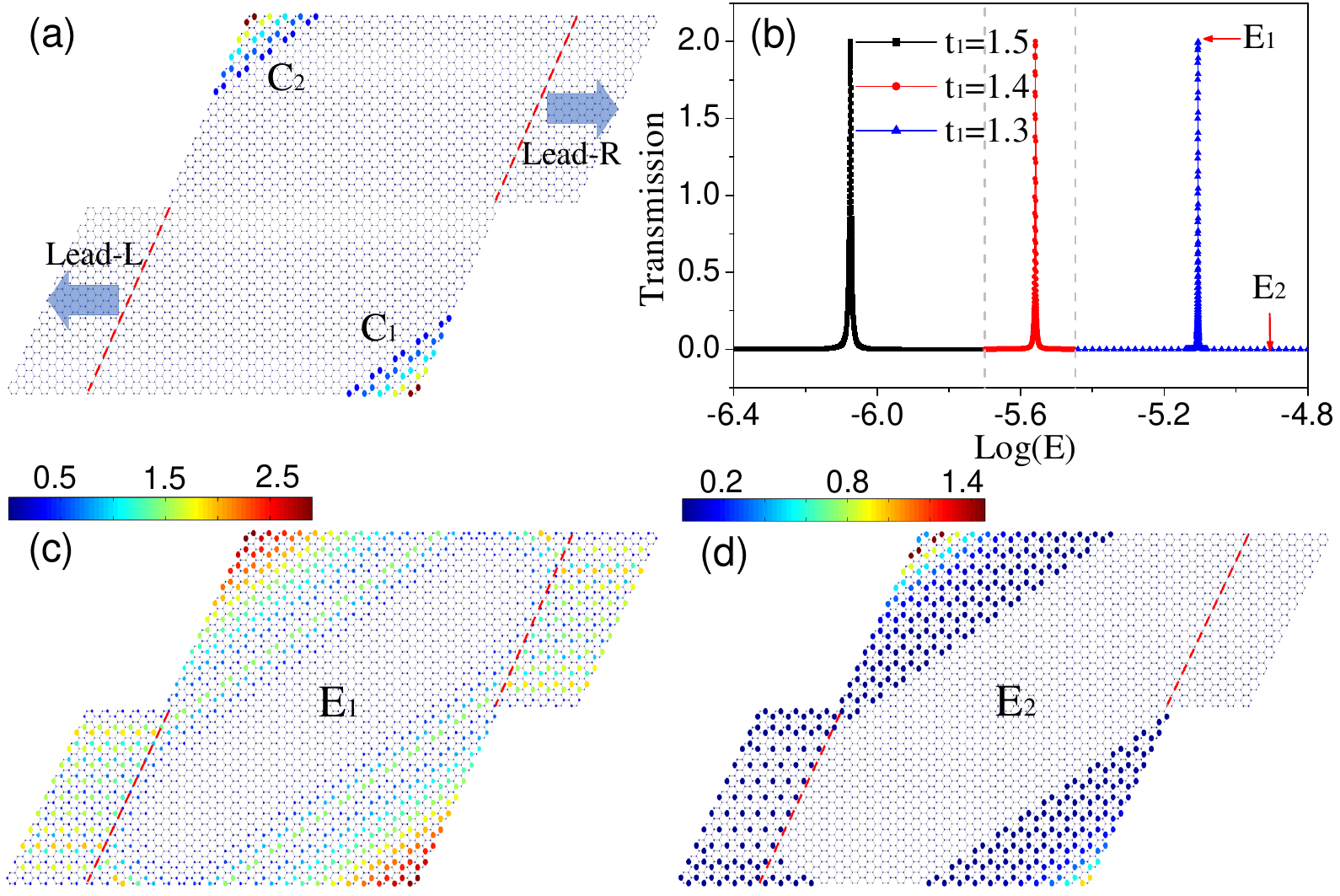}
\caption{(Color online) (a) Schematic of a two-terminal transport setup, where the diamond-shaped Kekul\'{e} flake is attached to narrow left and right leads. PLDOS at zero energy show that topological corner states locate at $C_{1}$ and $C_{2}$. (b) Transmission as a function of the Fermi energy for $t_{1}=1.3$, $1.4$, and $1.5$. The PLDOS corresponding to energies $E_{1}$ and $E_{2}$ are shown in (c) and (d), respectively. }\label{fig3}
\end{figure}

In the transport calculation, we consider a two-terminal setup as shown in Fig.~\ref{fig3}(a), where a diamond-shaped Kekul\'{e} lattice with armchair boundaries is connected to left and right conducting leads. The intracell hopping $t_{0}$ is set as the energy unit, and $t_{1} = 1.3t_{0}$ unless otherwise noted. Without loss of generality, the size of the diamond flake is fixed as $22*22$ unit cells in the following, while the width of the conducting leads is $11$ unit cells.

We first investigate transport features of topological corner states in a pristine Kekul\'{e} lattice, i.e., without local magnetization and electric potential. In the two-terminal transport setup in Fig.~\ref{fig3}(a), topological corner states are not destroyed by the connection to leads and still reside at corners $C_{1}$ and $C_{2}$ with obtuse angles.\cite{ref32} In Fig.~\ref{fig3}(b), the transmission as a function of the Fermi energy for different $t_{1}$ is plotted. It is found that multiple transmission peaks with $T=2$ appear when the energy $E$ sweeps the energy levels of corner states. $T=2$ indicates the existence of two spin degenerate conducting channels. As $t_{1}$ increases, the transmission peaks move towards the zero energy, which corresponds to more robust topological states. The narrow transmission peaks around zero energy result from the resonant tunneling assisted by topological corner states through the following process: incident electrons coming from the conducting leads transform the localized corner states into extended states across the whole scattering region, and hence conducing channel is formed in the insulating bulk.\cite{ref32,Xu2022} To demonstrate the dynamic process, we plot the PLDOS for different energies. In Fig.~\ref{fig3}(c) for $E = E_{1}$, the corner states at $C_{1}$ and $C_{2}$ are extended along the boundaries and connected to the conducting leads, making a bridge between the central region and leads. This phenomenon corresponds to the resonance tunneling process of topological corner states, which results in a perfect transmission peak with $T=2$. At $E = E_{2}$, since the energy is away from that of corner states, Fig.~\ref{fig3}(d) reveals that, electrons incident from the left lead can connect to the $C_1$ corner state but can not tunnel to the right lead; the $C_2$ corner state remains localized. As a result, the transmission at $E= E_{2}$ is $T \sim  0$.

\begin{figure}[tbp]
\includegraphics[width=\columnwidth]{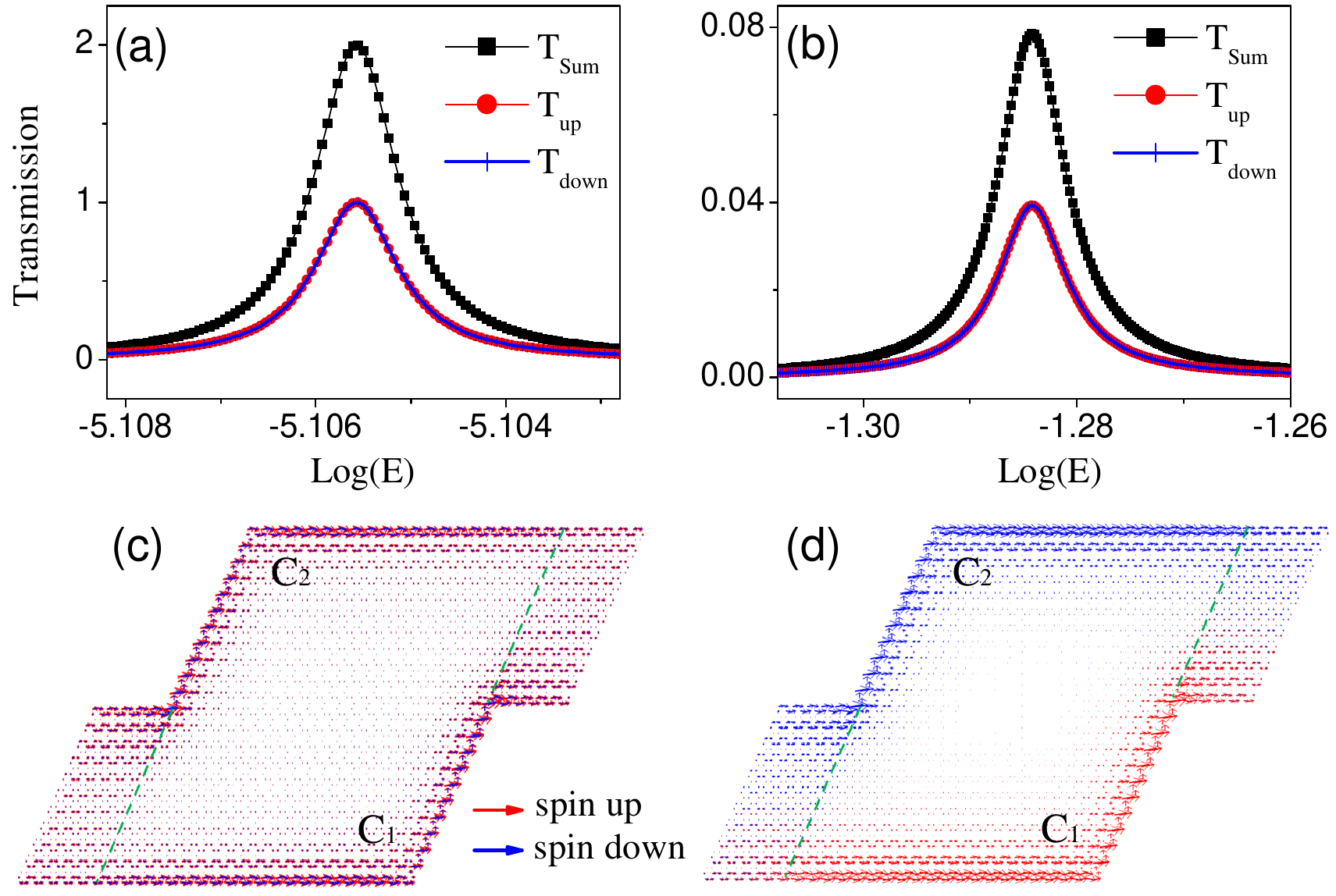}
\caption{(Color online) (a) Transmission versus the Fermi energy without local magnetization. (b) Transmission vs $E$ with opposite local magnetization applied on $C_1$ and $C_2$. The local current density distributions corresponding to the transmission peaks are plotted in (c) and (d), respectively. Green dash lines separate the central region and conducting leads. Parameters: $t_{1}=1.3$ and $\lambda=0.3$.}\label{fig4}
\end{figure}

Next, we study the effect of local magnetization on the transport behavior. The transmission spectrum from Fig.~\ref{fig3}(b) with $t_{1}=1.3$ is reproduced in Fig.~\ref{fig4}(a), where the transmission peak is enlarged for better comparison. Clearly, the spin-up and spin-down components contribute equally to the total transmission. When opposite local magnetization is applied on $C_1$ and $C_2$ corners (Fig.~\ref{fig2}(a)), $T$ as a function of $E$ is plotted in Fig.~\ref{fig4}(b). It is found that the existence of opposite local magnetization significantly reduces the transmission peak, leading to $T_{max} \sim 0.08$. As shown in Fig.~\ref{fig2}(b) and (c), the diamond-shaped lattice hosts spin-polarized topological corner states, and Fig.~\ref{fig4}(b) reveals that both of them still have equal contribution in $T_{Sum}$. In order to show the influence of local magnetization vividly, the local current density distribution of the two-terminal system is exhibited in Fig.~\ref{fig4}(c) and (d). Here, the local current vector is described by the arrow, where the size and orientation of the arrows denote the magnitude and direction of the local current density. The spin-up and spin-down local current vectors are highlighted in red and blue arrows, respectively. Fig.~\ref{fig4}(c) corresponds to the transmission peaks in Fig.~\ref{fig2}(a) without local magnetization. It is found that spin-degenerate current densities distribute mainly along the edges of the central region, connecting the $C_{1}$ and $C_{2}$ corner states with the conducting states in the leads, which coincides with the fact shown in Fig.~\ref{fig3}(c). In contrast, when local magnetization exists, Fig.~\ref{fig4}(d) demonstrates that the incident electrons with different spins are separated, and spin-up and spin-down scattering states propagate along the bottom and top boundaries of the central region, respectively. Spin polarization of topological corner states only allows incident electrons with the same spin to tunnel through one particular corner, which leads to the spatial separation of spin degree of freedom.

\begin{figure}[tbp]
\includegraphics[width=\columnwidth]{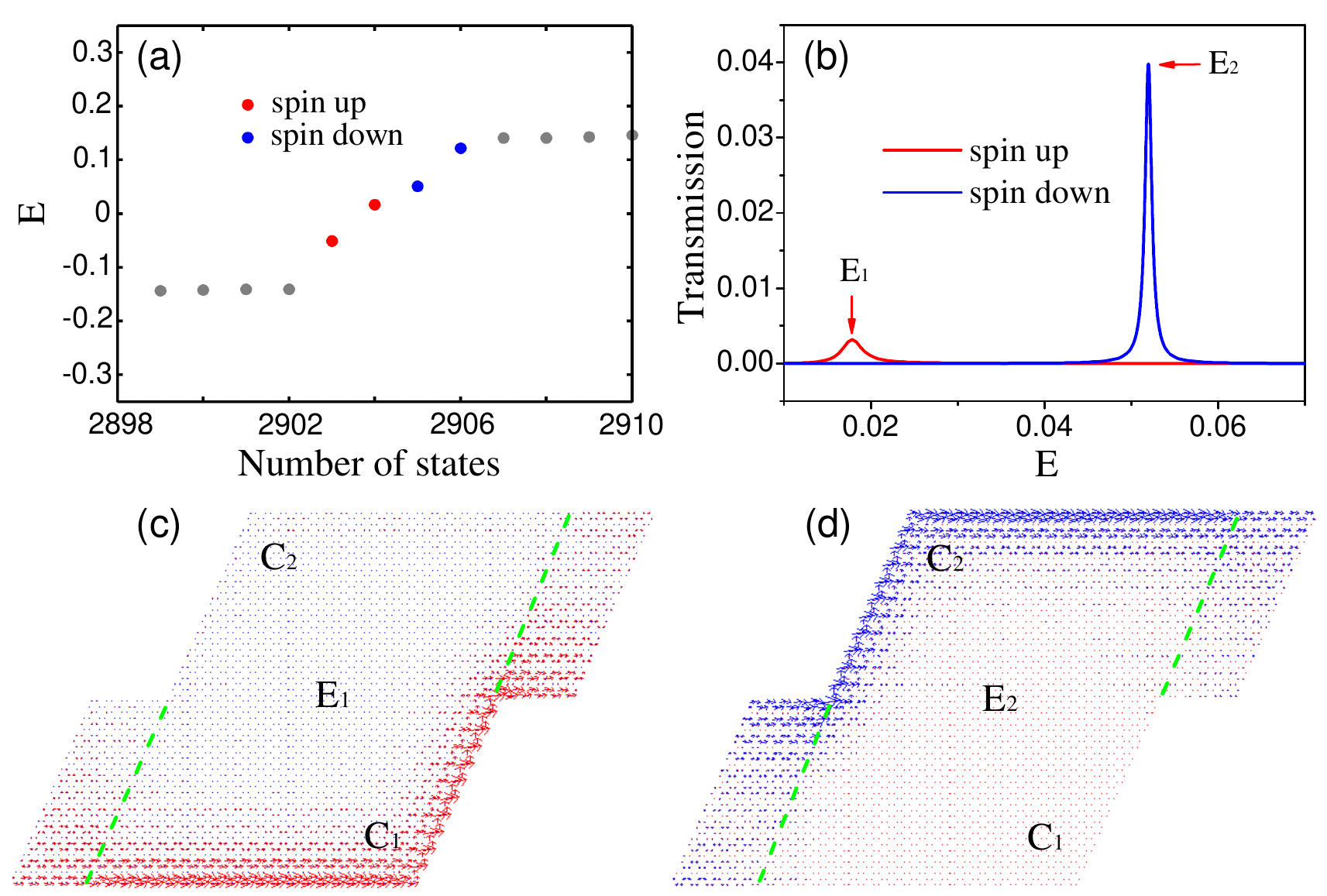}
\caption{(Color online) (a) Energy levels of the diamond-shaped Kekul\'{e} flake with the same local magnetization on both corners and an electric potential on the $C_1$ corner, as shown in Fig.~\ref{fig2}(d). The spin-up and spin-down corner states are highlighted in red and blue, respectively. (b) Spin-resolved transmission versus the Fermi energy. Local current density distributions corresponding to the transmission peaks at $E_{1}$ and $E_{2}$ are displayed in (c) and (d), respectively. Other parameters: $t_{1}=1.3$, $\lambda=0.3$, and $V=0.4$.}\label{fig5}
\end{figure}

In the following, we evaluate the interplay between local magnetization and electric potential, as well as their influence on transport properties. In Fig.~\ref{fig5}(a), we depict the energy levels of the isolated diamond-shaped Kekul\'{e} lattice with the same local magnetization and an electric potential on the $C_1$ corner, which is exactly the case shown in Fig.~\ref{fig2}(d). Different from the energy levels of a pristine lattice presented in Fig.~\ref{fig1}(c), Fig.~\ref{fig5}(a) shows that the spin degeneracy is lifted and the levels of spin-polarized corner states labeled by red and blue deviate from zero energy. Spin-resolved transmission functions in Fig.~\ref{fig5}(b) have two spin-dependent peaks, which correspond to the two adjacent levels with opposite spin polarizations in Fig.~\ref{fig5}(a)(red and blue dots in the middle). Introducing the local electric potential not only causes the separation of corner states in the energy spectrum. The local current density provides further transport details. We choose two typical energy points from Fig.~\ref{fig5}, $E_1$ and $E_2$, to observe their local current density vectors. It can be clearly seen from Fig.~\ref{fig5}(c): for $E_{1}$, the spin-up electrons incident from the left lead can flow into the right lead through the corner $C_{1}$, but spin-down electrons are forbidden to enter the central region. In Fig.~\ref{fig5}(d) for $E_{2}$, the system only allows the spin-down electrons to tunnel through the central scattering region via the corner $C_{2}$ and the spin-up conducting channel is closed. As a result, the coexistence of identical local magnetization and electric potential can result in spin-polarized currents for different energies, which means that the spin-polarized corner states can filter different spins by tuning the energy.

\begin{figure}[tbp]
\includegraphics[width=\columnwidth]{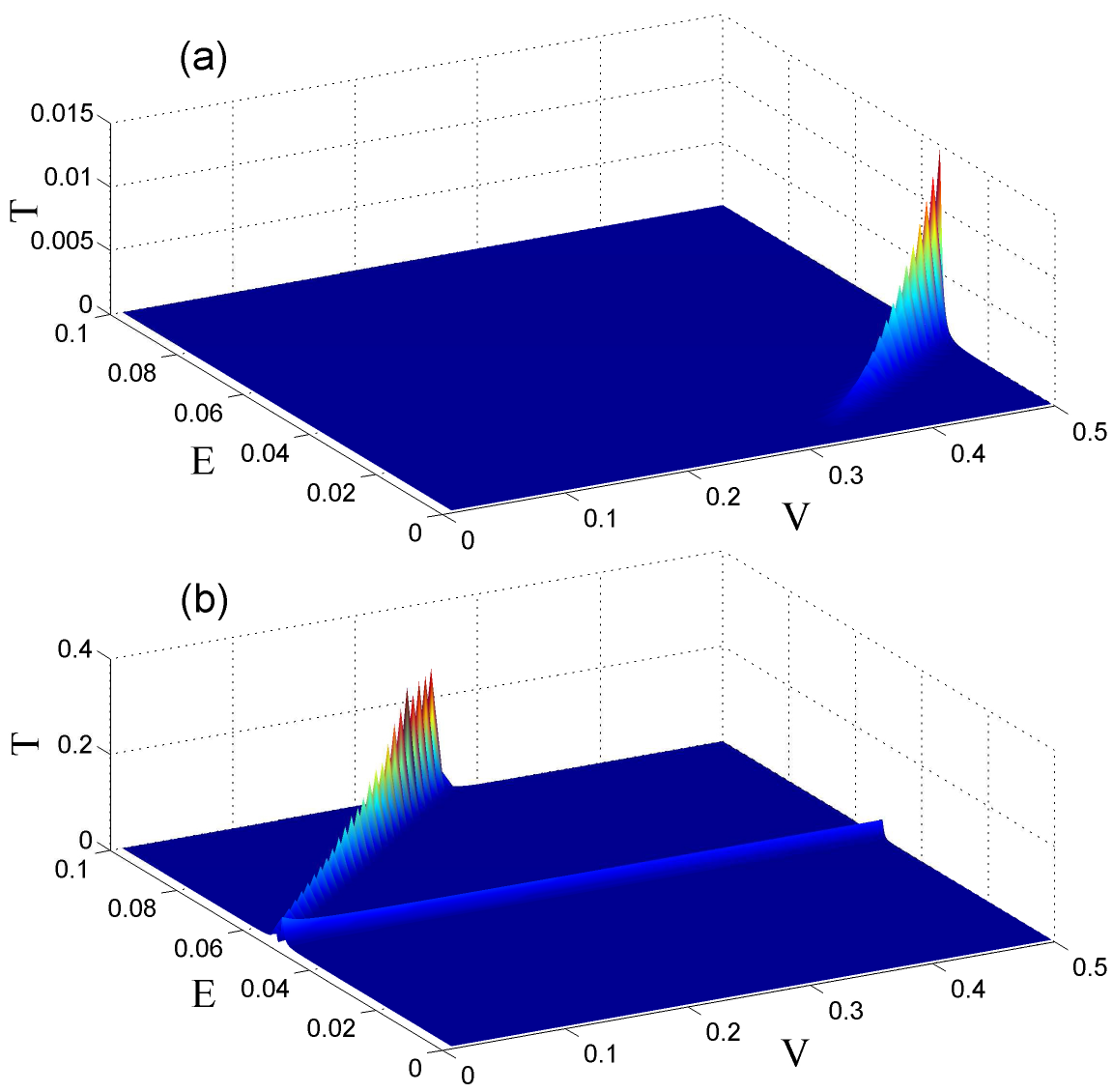}
\caption{(Color online) Transmission with respect to both Fermi energy $\rm E$ and electric potential $\rm V$ in the two-terminal setup. The spin-up and spin-down transmission diagrams are plotted in (a) and (b), respectively. The parameters are the same as those in Fig.~\ref{fig5}.}\label{fig6}
\end{figure}

To describe the influence of the electric potential on spin filtering, we provide spin-resolved transmission diagrams with respect to both the Fermi energy $E$ and the potential $V$. In Fig.~\ref{fig6}(a), the spin-up transmission is plotted. The results show that there is no conducting channel under small potential. However, when the potential exceeds the threshold $V=0.3$, the spin-up propagating channel is open, generating a spin-polarized current in the system. For the spin-down transmission, it is found from Fig.~\ref{fig6}(b) that a long ridge-like structure with constant transmission magnitude appears at $E=0.05$, which means this spin-down corner state robustly contributes to transport and is not affected by changing the electric potential. Moreover, another spin-down resonant channel can be observed for $E > 0.05$, where the transmission increases with the increasing of both $E$ and $V$. We deduce that these two spin-down channels originate from the spin-polarized corner states highlighted by blue circles in Fig.~\ref{fig5}(a). Our results demonstrate that, by varying $E$ and $V$, spin-polarized corner states provide the feasibility to effective spin filtering.

\section{Conclusions}

In summary, we have numerically investigated the spin-dependent transport properties of topological corner states in the Kekul\'{e} lattice. For a pristine system without local magnetization or electric potential, incoming electrons can tunnel through the central region via the corner states. Partial local density of states reveal that resonant conducting channels are formed through the corner states, resulting in perfect transmission peaks. The topological corner states can be spin-polarized with the application of local magnetization and electric potential at the corners. In the case of opposite local magnetization, when resonant tunneling occurs, incident electrons with spin-up and spin-down polarizations propagate along different corners, inducing the spatial separation of spin-polarized currents. When identical local magnetization and an electric potential are applied, the spin-polarized corner states can filter electrons of different spins by tuning the Fermi energy. Furthermore, the transmission diagram illustrates that various spin filtering configurations can be achieved by varying the Fermi energy and electric potential. Our results provide an effective way for spin manipulation via topological corner states.

$${\bf ACKNOWLEDGMENTS}$$

This work was supported by the National Natural Science Foundation of China (Grants No. 12304061, No. 12004102, and No. 12174262), the China Postdoctoral Science Foundation (Grant No. 2020M670836), the Natural Science Youth Foundation of Henan Province (Grant No. 232300421350). F. Xu also acknowledges the Shenzhen Science and Technology Program (JCYJ20220818100204010).

%---------------------------------------------------------------------------


\begin{thebibliography}{00}

%--------------Seebeck and Nernst Effect
\bibitem{ref1}
W. A. Benalcazar, B. A. Bernevig, and T. L. Hughes, Quantized electric multipole insulators, Science \textbf{357}, 61 (2017).

\bibitem{ref2}
F. Schindler, A. M. Cook, M. G. Vergniory, Z. Wang, S. S. P. Parkin, B. A. Bernevig, and T. Neupert, Higher-order topological insulators, Sci. Adv. \textbf{4}, eaat0346 (2018).

\bibitem{ref3}
C. W. Peterson, T. Li, W. A. Benalcazar, T. L. Hughes, and G. Bahl, A fractional corner anomaly reveals higher-order topology, Science \textbf{368}, 1114 (2020).

\bibitem{ref4}
B. Xie, H.-X. Wang, X. Zhang, P. Zhan, J.-H. Jiang, M. Lu and Y. Chen, Higher-order band topology, Nat. Rev. Phys. \textbf{3}, 520 (2021).

\bibitem{ref43}
Y. Ren, Z. Qiao, and Q. Niu, Engineering corner states from two-dimensional topological insulators, Phys. Rev. Lett. \textbf{124}, 166804 (2020).

\bibitem{ref44}
B. Wang, X. Zhou, H. Lin, and A. Bansil, Higher-order topological insulator phase in a modified Haldane model, Phys. Rev. Lett. \textbf{104}, 121108 (2021).

\bibitem{ref45}
H.-J. Lin, H.-P. Sun, T. Liu and P.-L. Zhao, Tuning three-dimensional higher-order topological insulators by surface state hybridization, Phys. Rev. B \textbf{108}, 165427 (2023).

\bibitem{Guo2023}
C.-A. Li, J. Sun, S.-B. Zhang, H. Guo, and B. Trauzettel, Klein-bottle quadrupole insulators and Dirac semimetals, Phys. Rev. B \textbf{108}, 235412 (2023).

\bibitem{ref6}
Z. Song, Z. Fang, and C. Fang, $(d-2)$-dimensional edge states of rotation symmetry protected topological states, Phys. Rev. Lett. \textbf{119}, 246402 (2017).

\bibitem{ref7}
X. Cheng, J. Chen, L. Zhang, L. Xiao, and S. Jia, Antichiral edge states and hinge states based on the Haldane model, Phys. Rev. B \textbf{104}, L081401 (2021).

\bibitem{ref8}
X.-L. Sheng, C. Chen, H. Liu, Z. Chen, Z.-M. Yu, Y. X. Zhao and S. A. Yang, Two-dimensional second-order topological insulator in graphdiyne, Phys. Rev. Lett. \textbf{123}, 256402 (2019).

\bibitem{Sheng2020}
C. Chen, Z. Song, J.-Z. Zhao, Z. Chen, Z.-M. Yu, X.-L. Sheng, and S. A. Yang, Universal Approach to Magnetic Second-Order Topological Insulator, Phys. Rev. Lett. \textbf{125}, 056402 (2020).

\bibitem{ref9}
Z. H. Li, P. Zhou, Q. H. Yan, X. Y. Peng, Z. S. Ma and L. Z. Sun, Second-order topological insulator in two-dimensional $\rm C_{2}N$ and its derivatives, Phys. Rev. B \textbf{106}, 085126 (2022).

\bibitem{ref10}
Y. Fang and J. Cano, Higher-order topological insulators in antiperovskites, Phys. Rev. B \textbf{101}, 245110 (2020).

\bibitem{ref46}
S. Regmi, M. M. Hosen, B. Ghosh, B. Singh, G. Dhakal, C. Sims, B. Wang, F. Kabir, K. Dimitri, Y. Liu, A. Agarwal, H. Lin, D. Kaczorowski, A. Bansil and M. Neupane, Temperature-dependent electronic structure in a higher-order topological insulator candidate $\rm EuIn_{2}As_{2}$, Phys. Rev. B \textbf{102}, 165153 (2020).

\bibitem{Sun2022} % 2nd-TI in 2D
Z. Li, P. Zhou, Q. Yan, X. Peng, Z. Ma, and L. Z. Sun, Second-order topological insulator in two-dimensional C$_2$N and its derivatives, Phys. Rev. B \textbf{106}, 085126 (2022).

\bibitem{ref11}
M. Ezawa, Higher-order topological insulators and semimetals on the breathing kagome and pyrochlore lattices, Phys. Rev. Lett. \textbf{120}, 026801 (2018).

\bibitem{ref12}
M. Ezawa, Higher-order topological electric circuits and topological corner resonance on the breathing kagome and pyrochlore lattices, Phys. Rev. B \textbf{98}, 201402(R) (2018).

\bibitem{ref13}
F. Schindler, Z. Wang, M. G. Vergniory, A. M. Cook, A. Murani, S. Sengupta, A. Y. Kasumov, R. Deblock, S. Jeon, I. Drozdov, H. Bouchiat, S. Gu\'{e}ron, A. Yadani, B. A. Bernevig,  and T. Neupert, Higher-order topology in bismuth, Nat. Phys. \textbf{14}, 918 (2018).

\bibitem{ref14}
L. Aggarwal, P. Zhu, and T. L. Hughes, and V. Madhavan, Evidence for higher order topology in Bi and $\rm Bi_{0.92}Sb_{0.08}$, Nat. Commun. \textbf{12}, 4420 (2021).

\bibitem{ref15}
M. J. Park, Y. Kim, G. Y. Cho, and S. Lee, Higher-order topological insulator in twisted bilayer graphene, Phys. Rev. Lett. \textbf{123}, 216803 (2019).

\bibitem{ref16}
B. Liu, L. Xian, H. Mu, G. Zhao, Z. Liu, A. Rubio and Z. F. Wang, Higher-order band topology in twisted moir\'{e} superlattice, Phys. Rev. Lett. \textbf{126}, 066401 (2021).

\bibitem{ref17}
Z. Wang, B. J. Wieder, J. Li, B. Yan, and B. A. Bernevig, Higher-Order Topology, Monopole Nodal Lines, and the Origin of Large Fermi Arcs in Transition Metal Dichalcogenides $\rm XTe_{2}$ ($X =Mo, W$), Phys. Rev. Lett. \textbf{123}, 186401 (2019).

\bibitem{ref18}
J. Zeng, H. Liu, H. Jiang, Q.-F. Sun and X. C. Xie, Multiorbital model reveals a second-order topological insulator in $\rm 1H$ transition metal dichalcogenides, Phys. Rev. B \textbf{104}, L161108 (2021).

\bibitem{ref19}
M. Serra-Garcia, V. Peri, R. S{\"u}sstrunk, O. R. Bilal, T. Larsen, L. G. Villanueva and S. D. Huber, Observation of a phononic quadrupole topological insulator, Nature \textbf{555}, 342 (2018).

\bibitem{ref20}
X.-D. Chen, W.-M. Deng, F.-L. Shi, F.-L. Zhao, M. Chen and J.-W. Dong, Direct observation of corner states in second-order topological photonic crystal slabs, Phys. Rev. Lett. \textbf{122}, 233902 (2019).

\bibitem{ref21}
S. Imhof, C. Berger, F. Bayer, J. Brehm, L. W. Molenkamp, T. Kiessling, F. Schindler, C. H. Lee, M. Greiter, T. Neupert and R. Thomale, Topolectrical-circuit realization of topological corner modes, Nat. Phys. \textbf{14}, 925 (2018).

\bibitem{ref24}
C. W. Peterson, W. A. Benalcazar, T. L. Hughes and G. Bahl, A quantized microwave quadrupole insulator with topologically protected corner states, Nature \textbf{555}, 7696 (2018).

\bibitem{ref48}
Z. Zhang, B. Hu, F. Liu, Y. Cheng, X. Liu, and J. Christensen, Pseudospin induced topological corner state at intersecting sonic lattices, Phys. Rev. B \textbf{101}, 220102(R) (2020).

\bibitem{ref23}
J. Wu, X. Huang, J. Lu, Y. Wu, W. Deng, F. Li and Z. Liu, Observation of corner states in second-order topological electric circuits, Phys. Rev. B \textbf{102}, 104109 (2020).

\bibitem{ref22}
X. Huang, J. Lu, Z. Yan, M. Yan, W. Deng, G. Chen and Z. Liu, Acoustic higher-order topology derived from first-order with built-in Zeeman-like fields, Sci. Bull. \textbf{67}, 488 (2022).

\bibitem{ref47}
S.-Q. Wu, Z.-K. Lin, B. Jiang, X. Zhou, Z. H. Hang, B. Hou and J.-H. Jiang, Higher-order Topological States in Acoustic Twisted Moir\'{e} Superlattices, Phys. Rev. Appl. \textbf{17}, 034061 (2022).

\bibitem{Sun2023} %
F. F. Huang, P. Zhou, W. Q. Li, S. D. He, R. Tan, Z. S. Ma, and L. Z. Sun, Phononic second-order topological phase in the C$_3$N compound, Phys. Rev. B \textbf{107}, 134104 (2023).

%\bibitem{ref49}
%S. Wang, H. Jia, X. Yang, P. Zhang, Y. Yang, Y. Yang, and X. Li, Realization of a full hierarchical topology in hexagonal bilayer
%acoustic crystals, Sci. China Phys. Mech. Astron. \textbf{66}, 104311 (2023).

%\bibitem{ref50}
%Y. Wang, B. Liang, and J. Cheng, The higher-order topological pumping explored in the 2D acoustic crystal, Sci. China Phys. Mech. Astron. \textbf{67}, 224311 (2024).

\bibitem{ref25}
Y. Wu, H. Jiang, J. Liu, H. Liu, and X. C. Xie, Non-abelian braiding of Dirac fermionic modes using topological corner states in higher-order topological insulator, Phys. Rev. Lett. \textbf{125}, 036801 (2020).

\bibitem{ref26}
R. Banerjee, S. Mandal, and T. C. H. Liew, Coupling between exciton-polariton corner modes through edge states, Phys. Rev. Lett. \textbf{124}, 063901 (2020).

\bibitem{ref27}
Y. Ota, F. Liu, R. Katsumi, K. Watanabe, K. Wakabayashi, Y. Arakawa, and S. Iwamoto, Photonic crystal nanocavity based on a topological corner state, Optica \textbf{6}, 786 (2019).

\bibitem{ref28}
X. Zhang, L. Liu, M.-H. Lu, and Y.-F. Chen, Valley-selective topological corner states in sonic crystals, Phys. Rev. Lett. \textbf{126}, 156401 (2021).

\bibitem{ref29}
F. Liu, H.-Y. Deng, and K. Wakabayashi, Helical topological edge states in a quadrupole phase, Phys. Rev. Lett. \textbf{122}, 086804 (2019).

\bibitem{Duan2017}
Y. Liu, C.-S. Lian, Y. Li, Y. Xu, and W. Duan, Pseudospins and Topological Effects of Phonons in a Kekul\'{e} Lattice, Phys. Rev. Lett. \textbf{119}, 255901 (2017).

\bibitem{ref30}
F. Zangeneh-Nejad and R. Fleury, Nonlinear second-order topological insulators, Phys. Rev. Lett. \textbf{123}, 053902 (2019).

\bibitem{ref31}
Y. Zhou and R. Wu, Remote control of spin polarization of topological corner states, Phys. Rev. B \textbf{107}, 035412 (2023).

\bibitem{Hasan23}
M. D. Hossain, Q. Zhang, Z. Wang, N. Dhale, W. Liu, M. Litskevich, B. Casas, N. Shumiya, J.-X. Yin, T. A. Cochran, Y. Li, Y.-X. Jiang, Y. Zhang, G. Cheng, Z.-J. Cheng, X. P. Yang, N. Yao, T. Neupert, L. Balicas, Y. Yao, B. Lv, M. Z. Hasan, Quantum transport response of topological hinge modes, Nat. Phys. \textbf{20}, 776-782 (2023).

\bibitem{ref32}
K.-T. Wang, Y. Ren, F. Xu, Y. Wei, and J. Wang, Transport induced dimer state from topological corner states, Sci. China Phys. Mech. Astron. \textbf{64}, 257811 (2021).

\bibitem{Xu2022}
K.-T. Wang, F. Xu, B. Wang, Y. Yu, Y. Wei, Transport features of topological corner states in honeycomb lattice with multihollow structure, Front. Phys. \textbf{17}, 43501 (2022).

\bibitem{ref5}
Y. Long, M. Wei, F. Xu, and J. Wang, Bulk-boundary-transport correspondence of the second-order topological insulators, Sci. China Phys. Mech. Astron. \textbf{66}, 127811 (2023).

\bibitem{ref51}
J. Poata, F. Taddei, and M. Governale, Topological nano-switches in higher-order topological insulators, New J. Phys. \textbf{26}, 053038 (2024).

\bibitem{ref33}
C.-Y. Hou, C. Chamon and C. Mudry, Electron Fractionalization in Two-Dimensional Graphenelike Structures, Phys. Rev. Lett. \textbf{98}, 186809 (2007).

\bibitem{ref34}
E. Lee, A. Furusaki and B.-J. Yang, Fractional charge bound to a vortex in two-dimensional topological crystalline insulators, Phys. Rev. B \textbf{101}, 241109(R) (2020).

\bibitem{note0}
For the armchair Kekul\'{e} ribbon with finite width, there is a small finite-size gap in its band structure. To acheive a conducting ribbon with the width of $22$ unit cells, we set $t_{1} = 1.07~t_{0}$ in the calculation. The results are shown in Fig.~\ref{fig1}(b) as the gray background, which is very similar to those of a graphene armchair ribbon with zero gap.

\bibitem{ref35}
S. Datta, Electronic Transport in Mesoscopic System (Cambridge University Press, Cambridge, UK, 1995).

\bibitem{ref36}
D. H. Lee and J. D. Joannopoulos, Simple scheme for surface-band calculations. $\rm \uppercase\expandafter{\romannumeral1}$, Phys. Rev. B \textbf{23}, 4988 (1981).

\bibitem{ref37}
D. H. Lee and J. D. Joannopoulos, Simple scheme for surface-band calculations. $\rm \uppercase\expandafter{\romannumeral2}$. The Green's function, Phys. Rev. B \textbf{23}, 4997 (1981).

\bibitem{ref38}
V. Gasparian, T. Christen, and M. B{\"u}ttiker, Partial densities of states, scattering matrices, and Green's functions, Phys. Rev. A \textbf{54}, 4022 (1996).

\bibitem{ref39}
T. Gramespacher and M. B{\"u}ttiker, Nanoscopic tunneling contacts on mesoscopic multiprobe conductors, Phys. Rev. B \textbf{56}, 13026 (1997).

\bibitem{ref40}
O. Ly, R. A. Jalabert, S. Tomsovic, and D. Weinmann, Partial local density of states from scanning gate microscopy, Phys. Rev. B \textbf{96}, 125439 (2017).

\bibitem{ref41}
H. Jiang, L. Wang, Q.-F. Sun, and X. C. Xie, Numerical study of the topological Anderson insulator in HgTe/CdTe quantum wells, Phys. Rev. B \textbf{80}, 165316 (2009).

\bibitem{ref42}
Y. Xing, L. Zhang, and J. Wang, Topological Anderson insulator phenomena, Phys. Rev. B \textbf{84}, 035110 (2011).

%\bibitem{Poata2024}
%J. Poata, F. Taddei, and M. Governale, Topological nano-switches in higher-order topological insulators, New J. Phys. \textbf{26}, 053038 (2024).

%%add reference

%\bibitem{ref44}
%B. Wang, X. Zhou, H. Lin and A. Bansil, Higher-order topological insulator phase in a modified Haldane model, Phys. Rev. Lett. \textbf{104}, L121108 (2021).

%\bibitem{ref45}
%H.-J. Lin, H.-P. Sun, T. Liu and P.-L. Zhao, Tuning three-dimensional higher-order topological insulators by surface state hybridization, Phys. Rev. B \textbf{108}, 165427 (2023).

%\bibitem{ref46}
%S. Regmi, M. M. Hosen, B. Ghosh, B. Singh, G. Dhakal, C. Sims, B. Wang, F. Kabir, K. Dimitri, Y. Liu, A. Agarwal, H. Lin, D. Kaczorowski, A. Bansil and M. Neupane, Temperature-dependent electronic structure in a higher-order topological insulator candidate $\rm EuIn_{2}As_{2}$, Phys. Rev. B \textbf{102}, 165153 (2020).

%\bibitem{ref47}
%S.-Q. Wu, Z.-K. Lin, B. Jiang, X. Zhou, Z. H. Hang, B. Hou and J.-H. Jiang, Higher-order Topological States in Acoustic Twisted Moir\'{e} Superlattices, Phys. Rev. Appl. \textbf{17}, 034061 (2022).

%\bibitem{ref48}
%Z. Zhang, B. Hu, F. Liu, Y. Cheng, X. Liu, and J. Christensen, Pseudospin induced topological corner state at intersecting sonic lattices, Phys. Rev. B \textbf{101}, 220102(R) (2020).


\end{thebibliography}
\end{document}